%% file: main.tex
\documentclass{jaa}
\usepackage{natbib}
\usepackage[colorlinks=true,citecolor=blue]{hyperref}
\usepackage{amsmath}
\usepackage{graphicx}

\def\deg{\ifmmode^\circ\else$^\circ$\fi}

\pdfoutput=1

\begin{document}\sloppy

%%paper title
%%For line breaks \\ can be used within title
\title{Peeping into recent star formation history of the Magellanic Cloudss}
\author{Y.~C. Joshi\textsuperscript{1}, A. Panchal\textsuperscript{1}}
\affilOne{\textsuperscript{1}Aryabhatta Research Institute of Observational Sciences (ARIES) Manora Peak, Nainital-263002, India\\}

%%author names are separated by comma (,)
%%use \and before the last author name
%%use a * along with the number separated by comma
%% for the  author for correspondence
%%\textsuperscript{number} is used for affiliation
%%\affilOne, \affilTwo etc., upto \affilTwentyfive is possible
%%Please note the first letter after \affil is capitalised in the command
%%
%
%\author{AUTHOR1\textsuperscript{1}, AUTHOR2\textsuperscript{1} and AUTHOR3\textsuperscript{2,*}}
%\affilOne{\textsuperscript{1}Department of P, University X, Place Pincode, Country.\\}
%\affilTwo{\textsuperscript{2}Department of Q, University Z, Place Pincode, Country.}
%
%%escape two column mode for title, affiliation and abstract
%%by giving \twocolumn command as shown

\twocolumn[{

\maketitle

%%include \corres to print the corresponding author Email id
\corres{yogesh@aries.res.in}

%%include \msinfo for
%%manuscript information such as
%%received, revised and accepted dates
%%
\msinfo{***}{***}

%%abstract
\begin{abstract}
We study the distribution of Fundamental-mode Cepheids in the Magellanic Clouds as a function of their positions and ages using the data from the OGLE~IV survey. The ages of the Cepheids are determined through well known period - age relations for the LMC and SMC Cepheids which are used to understand the star formation scenario in the Magellanic Clouds. The age distributions of the Cepheids in LMC and SMC show prominent peaks around $158^{+46}_{-35}$ Myr and $219^{+63}_{-49}$ Myr, respectively. This indicates that a major star formation event took place in the Magellanic Clouds about 200\,Myr ago. It is believed that this episode of enhanced star formation might have been triggered by a close encounter between the two components of the Magellanic Clouds or due to a possible tidal interaction between the Magellanic Clouds and Milky Way galaxy during one of its pericentric passages around the Milky Way. Cepheids are found to be asymmetrically distributed in both the LMC and SMC in an elongated manner. A high spatial density clumpy structure is found to be located towards the eastern side of the LMC and the south-west direction of the SMC from their respective galactic centers.
\end{abstract}

%\keywords{Galaxies: active---galaxies: individual (NGC 4941)---X-rays:galaxies}

}]

%%close the twocolumn escape here

%%include \doinum{number}for the DOI number in the header
%%include \volnum{number} for the volume number in the header
%%include \year{yyyy} for  year of publication in the header
%%include \pgrange{num--num} page range of article in the header
%%include \artcitid{num} for the article citation id
%%include \lp to print last page of the article
%%include \setcounter{page}{pagenum} for the exact starting page of the article

\doinum{12.3456/s78910-011-012-3}
\artcitid{\#\#\#\#}
\volnum{000}
\year{0000}
\pgrange{1--}
\setcounter{page}{1}
\lp{1}

\section{Introduction}\label{sec:intro}
The Magellanic Clouds is among one of the most studied galaxy in the Universe due to its close proximity, favourable viewing angle, and star formation activities \citep{2005MNRAS.357..304H, 2006ApJ...642..834K}. Star formation can be triggered by several mechanisms like a turbulent interstellar medium, self-induced gravitational collapse of the molecular cloud, tidal shocking, or cloud-cloud interactions \citep[e.g.,][]{2012ApJ...750...36D, 2018MNRAS.473.3131M, 2019ApJ...874...78Z}. In recent times, the star formation activity in Magellanic Clouds has been studied through scanning of various stellar populations, e.g. star clusters \citep{2010A&A...517A..50G, 2022MNRAS.509.3462P}, Cepheid variables \citep{2015A&A...573A.135S, 2016AcA....66..149J, 2019A&A...628A..51J}, RR Lyrae variables \citep{2009A&A...503L...9S, 2013MNRAS.431.1565W, 2021MNRAS.504....1C, 2022MNRAS.512..563R}, red clump stars \citep{2009AJ....138....1K, 2013A&A...552A.144S, 2021ApJS..252...23S}, among others. These studies unanimously imply that the episodic star formation events have taken place in the Magellanic Clouds, most likely due to repeated interaction between the two components of the Magellanic Clouds and/or between the Magellanic Clouds and Milky Way Galaxy \citep{2010A&A...517A..50G, 2011A&A...535A.115I, 2014MNRAS.438.1067M, 2014MNRAS.445.2214R, 2016RAA....16...61J, 2018MNRAS.478.5017R, 2019A&A...628A..51J}.

Classical Cepheids have widely been used to probe the history of star formation in the Magellanic Clouds because these sources are intrinsically bright, easily observable, and ubiquitous in both Large Magellanic Clouds (LMC) and Small Magellanic Clouds (SMC) \citep{2016AcA....66..149J, 2019MNRAS.489.3725D}. The use of Classical Cepheids as tracers of young stellar populations comes from the fact that they obey the period-luminosity and period-age relations \citep{1998MNRAS.299..588E, 2005ApJ...621..966B, 2014NewA...28...27J}. They are thus ideal objects to understand the star formation activity in the past 30 - 600 Myrs of the galaxies as typical life of the Classical Cepheids spans in this age range. This led to their immense uses in tracing young stellar populations and star-forming regions in the extra-galactic systems \citep{1996ApJ...466..802E}.

In recent times, Optical Gravitational Lensing Experiment (OGLE) survey\footnote{http://ogle.astrouw.edu.pl/} has revolutionized the field by producing thousands of Cepheid light curves and derived parameters like periods, magnitudes, and amplitudes \citep{2017AcA....67..103S}. In this paper, we aim to understand the age distributions of the Cepheids in the LMC and SMC to probe the star formation history of the Magellanic Clouds. The paper is organized as follows: details of the data are given in Section~\ref{data} The period and age relations are discussed in Section~\ref{pa} The age, spatial and spatio-temporal distributions of Cepheids in the LMC and SMC are studied in Sections~\ref{sfh_l} and \ref{sfh_s} The detailed discussion of star formation in Magellanic Clouds is discussed in Section~\ref{discuss} We give summary of our work in Section~\ref{conc}.
\section{Data}\label{data}
The OGLE survey provides photometric data in the $V$ and $I$-bands for stars in a 40 square degree region of the LMC using the 1.3-m Warsaw telescope at the Las Campanas Observatory, Chile \citep{2015AcA....65..233S, 2017AcA....67..103S}. In the OGLE-IV survey, a catalogue of 2476 Fundamental-mode Cepheids in the LMC and 2753 Fundamental-mode Cepheids in SMC are made available on their site which was identified from their light curves and location in the period-luminosity diagram. The catalogue contains coordinates of the Cepheids, their mean $V$ and $I$ magnitudes, amplitudes, periods, and classification as well as cross-identification with other catalogues. Although there are other higher-mode or multi-mode Cepheids reported in the OGLE survey, their reliable age estimation is rather difficult, unlike the Fundamental-mode Cepheids. Unlike our earlier study in \citet{2019A&A...628A..51J} where we used both Fundamental-mode and First overtone Cepheids in the analysis, we here have considered only Fundamental-mode Cepheids in the LMC and SMC. This is to avoid the age estimation of the First overtone Cepheids through conversion of First overtone pulsation period to corresponding fundamental period by just supplying a single conversion factor which may not be true for all the pulsating Cepheids. For all the Fundamental-mode Cepheids (here onward we refer to Fundamental-mode Cepheids as Cepheids for simplicity), we also converted ($RA$, $DEC$) coordinates to corresponding ($X$, $Y$) coordinates using the relations given by \cite{1967pras.book.....V}.
\section{Period-Age relations}\label{pa}
The ages of the Cepheids are used to understand the star formation scenario in the Magellanic Clouds, however, there is no direct way of estimating of ages of Cepheids. The period-luminosity and mass-luminosity relations in classical Cepheids suggest that longer-period Cepheids have higher luminosities hence more massive which means these stellar populations are relatively younger in comparison to shorter-period Cepheids. This suggests that the period and age of Cepheids have an obvious connection. Since the period of a large number of Cepheids has been estimated with very high accuracy in the OGLE-IV survey having an error less than 0.001\%, we used the period-age relations known for Cepheids in the LMC and SMC to determine their respective ages. To estimate the ages of Cepheids from their pulsation periods, many authors have provided semi-empirical relations in the past \citep{1997A&AS..126..401M, 1998MNRAS.299..588E}. In our previous study in \citet{2014NewA...28...27J}, we used 74 LMC Cepheids found in 25 different open clusters to derive their period-age relation. For this purpose, we considered the mean period of Cepheids reported by \citet{2003ARep...47.1000E} and their corresponding cluster ages from \citet{2010MNRAS.403.1491P} to draw an empirical Period-Age (PA) relation in the LMC for the Cepheids in the LMC which is given as
\begin{equation}
\log ({\rm t}) = 8.60(\pm0.07) - 0.77(\pm0.08)~\log ({\rm P})
\label{eqno1} 
\end{equation}
where age denoted by $t$ is in years and P is the period given in days. For the SMC Cepheids, the following PA relation given by \cite{2005ApJ...621..966B} has been used to estimate the age of the Cepheids.
\begin{equation}
\log ({\rm t}) = 8.49(\pm0.09) - 0.79(\pm0.01)~\log ({\rm P})
\label{eqno2} 
\end{equation}
We used the above PA relations to determine the age of each Cepheid variable in the LMC and SMC. We found a typical average error of $\sim$40 Myr in the age estimations through these PA relations. Considering an insignificant age-metallicity gradient in the Magellanic Clouds \citep{2013AJ....145...17P} in comparison to the error associated with the age estimation of the Cepheids which are relatively younger populations, we have not taken into account any metallicity dependency on the above PA relations.
\section{Star formation history in the Large Magellanic Cloudss}\label{sfh_l}
\subsection{Age distribution of Pop I Cepheids}\label{age_l}
\begin{figure}
\centering
\includegraphics[width=8.0cm, height=8.0cm]{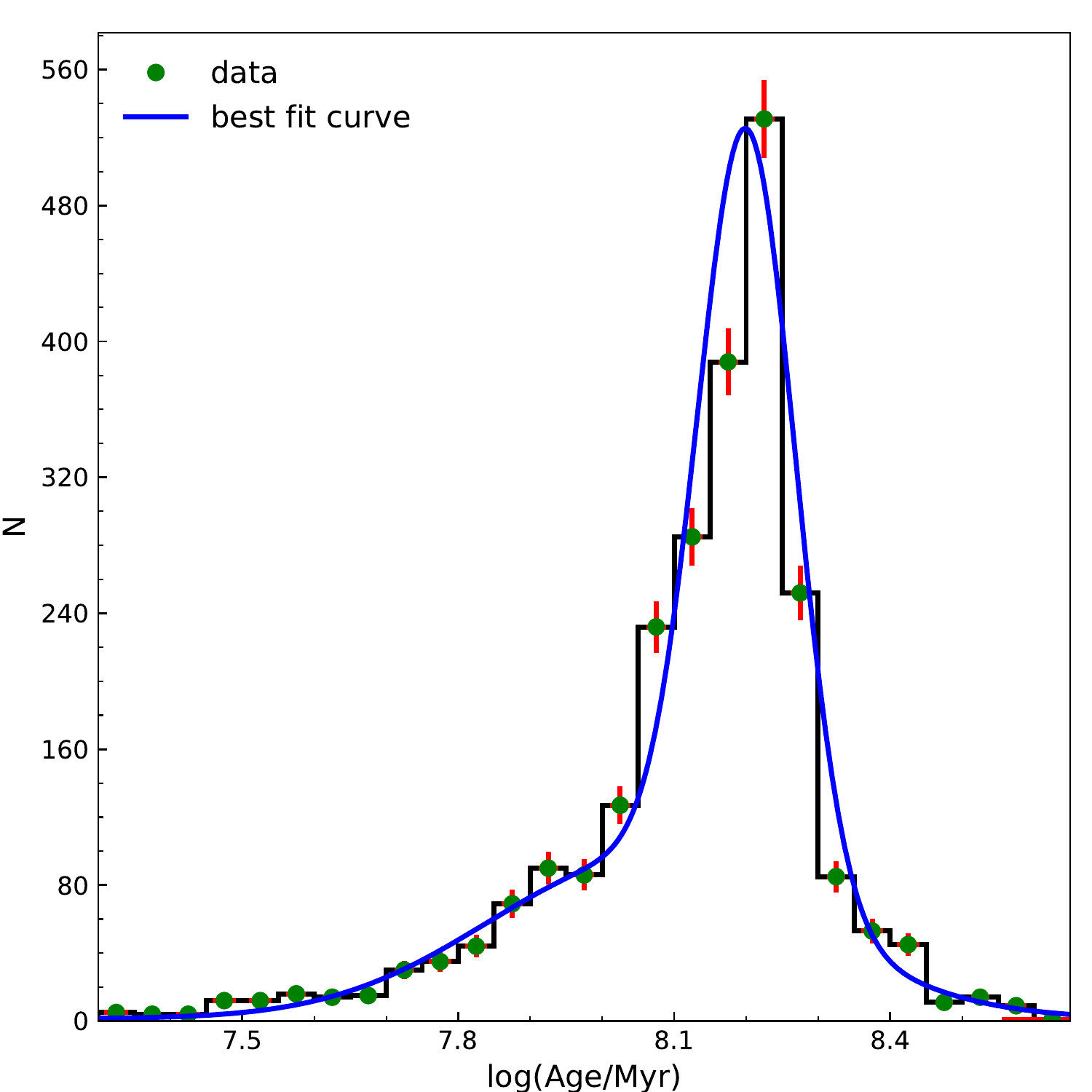}
\caption{Age distribution of Cepheids in the LMC. The best-fit Gaussian profile is shown by a thick continuous line.}
\label{fig01}
\end{figure}

We estimated the age of each Cepheid in our sample of LMC Cepheids using the equation~\ref{eqno1}. The resulting age distribution of Cepheids is shown in Figure~\ref{fig01} for a bin width of $\Delta(log(Age)) = 0.05$. The standard deviation of the log (Age) of all the Cepheids in a bin is used as x-axis error bar for that specific bin. Whereas there is only one Cepheid in the bin, the error bar represents the errors derived from equation~\ref{eqno1}. The y-axis error bar for each bin corresponds to the poison noise of the number of Cepheids. The LMC age distribution rises gradually but shows a steep fall. A single Gaussian fit as well as a double Gaussian fit are tested for this distribution. The single Gaussian fit to the distribution peaks at $\log ({\rm t/yr}) = 8.16\pm0.10$ corresponding to an age of $145^{+37}_{-30}$ Myr. The double Gaussian fit shows peaks at $\log ({\rm t/yr}) = 8.06\pm0.11$ and $8.20\pm0.11$ corresponding to age of $115^{+33}_{-26}$ Myr and $158^{+46}_{-35}$ Myr, respectively. Though errors in the fitting coefficients are very small but quoted uncertainty in the peak age represents a combined error which arises from the errors on the coefficients in equation~\ref{eqno1} and statistical error in the Gaussian fit. We compared the observed distribution with fitted distribution using Kolmogorov–Smirnov (K-S) test for both single and double Gaussian fits. The K-S test statistics and p-values are listed in Table~\ref{fit_test}. A lower $\chi^{2}$ and higher q-vale from K-S test favors the double Gaussian trend of the LMC log(Age) distribution. However, secondary peak in the age distribution of Cepheids in the LMC is not prominent enough to reveal any major star formation at this epoch.
%
%---------------------------- Fig01 -----------------------------------
%------------------------------------------------------------------------
%

\subsection{Spatial distribution}\label{spatial_l}
\begin{figure}
\centering
\vspace{-0.9cm}
\includegraphics[width=9.5cm, height=9.0cm]{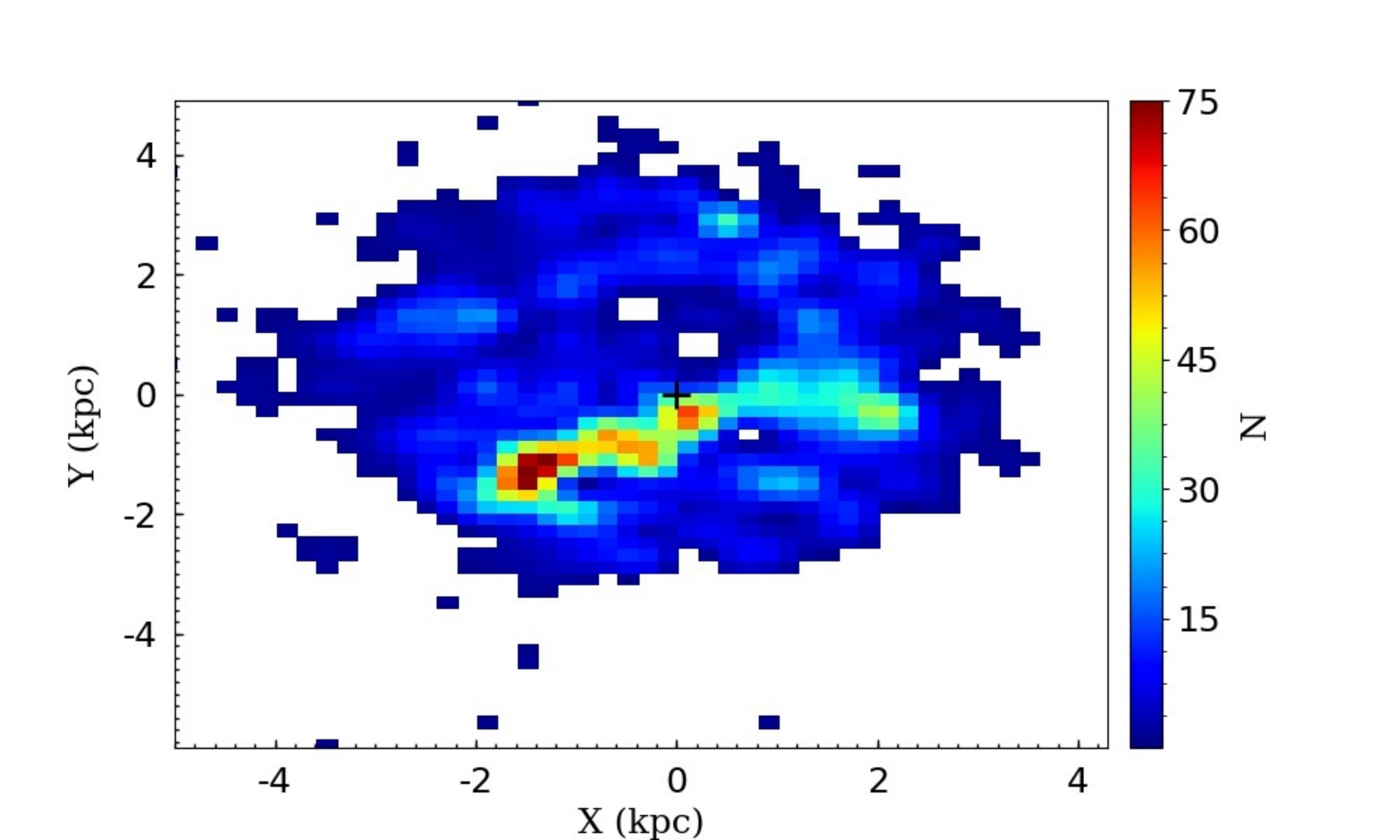}
\caption{Spatial heat map of the LMC where colour represents the number of Cepheids found in different segments. North is up and east is to the left. The position of the known center of the LMC is shown by a plus symbol.}
\label{fig02}
\end{figure}

In many previous studies, the distribution of Cepheids has been used to probe the structure of this dwarf cloud \citep[e.g.][]{2014NewA...28...27J, 2016ApJ...832..176I}. To study the spatial distribution of Cepheids in the LMC, we divided the whole sample of Cepheids into smaller segments with a dimension of $0.4\times0.4$ kpc square that results in a spatial resolution of $\approx$ 1.2 deg$^2$. The bin size was chosen to make an optimum number of segments without degrading much in the spatial resolution. Our selection resulted in 286 segments in the LMC. In Figure~\ref{fig02}, we illustrate the heat map of the spatial distribution of Cepheids where colour code represents the number density of Cepheids in each segment. A clumpy structure in the map is visible close to $(X,Y) = (-1.58,-1.18)$ which is towards the south-east direction from the known center of the LMC at $(X,Y) = (0,0)$ that corresponds to ($\alpha, \delta$) = ($5^h19^m38^s, -69^0 27^{'} 5^{''}.2)$), and shown by a plus sign in Figure~\ref{fig02}. It is evident from Figure~\ref{fig02} that the center of LMC is not quite coincident with the dense region of the Cepheids which suggests that the enhanced star formation is not exactly taking place in the central part of the LMC.
%
%---------------------------- Fig02 -----------------------------------
We also notice a bar-like elongated composition in the east-west direction of the LMC which is conspicuous by the relatively larger density of Cepheids. The bar in the LMC is a well-observed structure in many previous studies inferred through different stellar populations including Cepheids \citep{2018MNRAS.478.2526D, 2019A&A...628A..51J}, RR Lyrae stars \citep{2009A&A...503L...9S, 2021MNRAS.504....1C, 2022MNRAS.512..563R}, and Red Clump stars \citep{2013A&A...552A.144S, 2021ApJS..252...23S}. Despite evidence of bar structure in the LMC, Cepheids are not uniformly distributed in the bar and there are few structures within the bar where clumpy regions are more prominent, particularly in the south-east region that is far off from the LMC center. This sort of patchy star formation within the bar has also been noticed by \citet{2021MNRAS.508..245M}. However, \citet{2018MNRAS.473L..16M} ruled out any such significant spatial variations of the star formation across the LMC bar. A poor spatial density of Cepheids is also quite conspicuous in the northern arm of the LMC as evident in Figure~\ref{fig02}.
%
%---------------------------- Fig03 -----------------------------------
\begin{figure}
\centering
\vspace{-0.9cm}
\includegraphics[width=9.5cm, height=8cm]{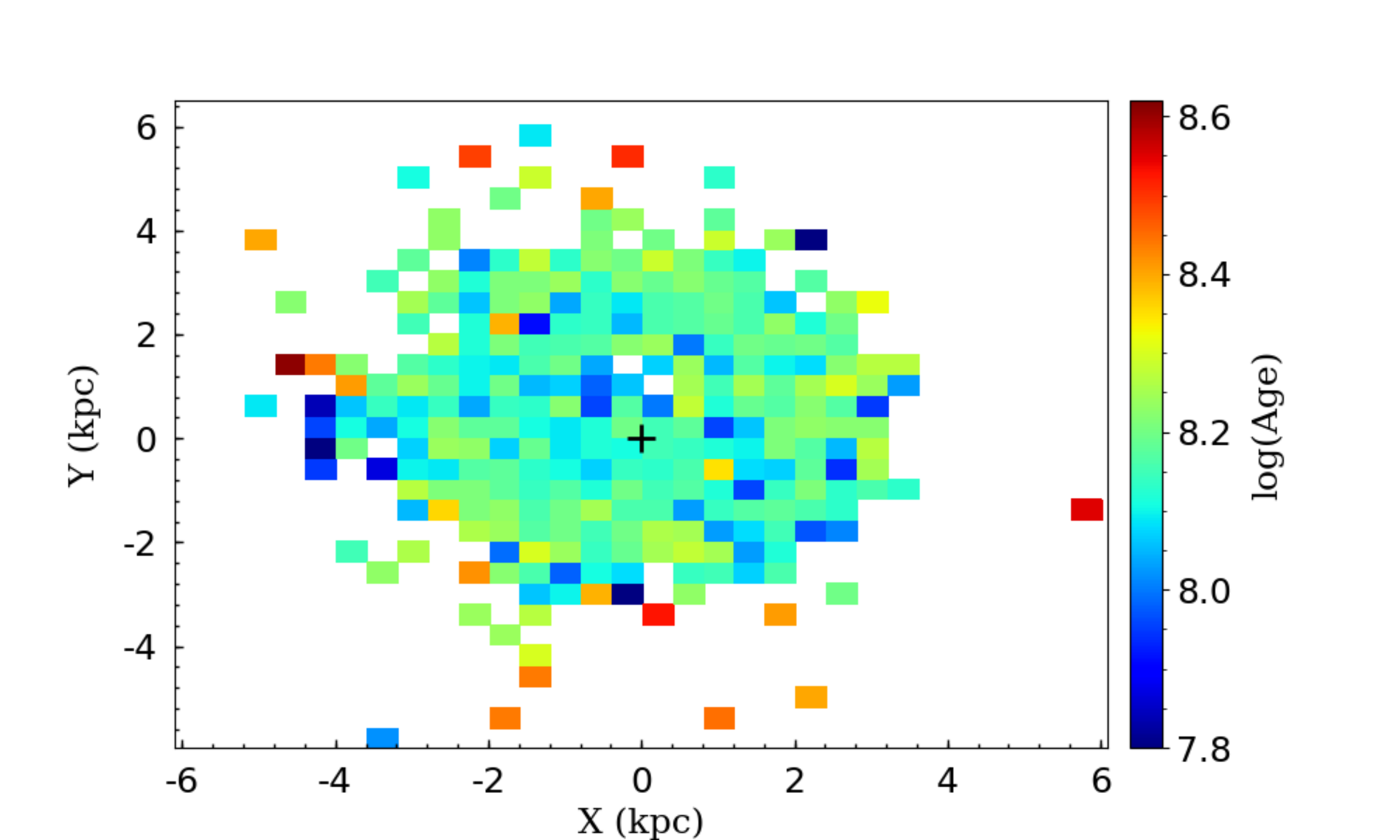}
\caption{Spatial distributions of the Cepheids in the 286 segments of the LMC as a function of their mean $\log ({\rm t/yr})$. North is up and east is to the left. The position of the known center of the LMC is shown by a plus symbol.}
\label{fig03}
\end{figure}
%------------------------------------------------------------------------
%
\subsection{Spatio-temporal distribution}\label{temporal_l}
To understand spatio-temporal distribution of Cepheids in 286 segments of the LMC, we illustrate the heat map of the spatial distribution of Cepheids as a function of their mean log(Age) in Figure~\ref{fig03}. Most of the Cepheids in the inner region have almost uniform age distribution, where mostly intermediate age Cepheids in addition to some young age Cepheids are seen. However, there is some sort of inhomogeneity towards peripheral regions where many pockets of old-age Cepheids are observed. This gradual change in the age distribution of Cepheids from the inner region to the outer region suggests an outside-in quenching of star formation in the LMC disk. A similar structure in age gradient was also noticed by \citet{2014MNRAS.438.1067M} in the LMC. It is believed that this stellar quenching may be resulted due to gas depletion at the time of star formation or ram-pressure stripping \citep{2014MNRAS.438.1067M}. The outside-in quenching of star formation in the outer LMC disk is believed to be associated with the variation of the size of the HI disk as a consequence of gas depletion due to enhance star formation or ram-pressure stripping, or of the compression of the gas disk as ram pressure from the Milky Way halo acted on the LMC interstellar medium \citep{2014MNRAS.438.1067M}.
%
%Table 01
\input{tab01.tex}
\section{Star formation history in the Small Magellanic Cloudss}\label{sfh_s}
\subsection{Age distribution}\label{age_s}
\begin{figure}
\centering
\includegraphics[width=8.0cm, height=8.0cm]{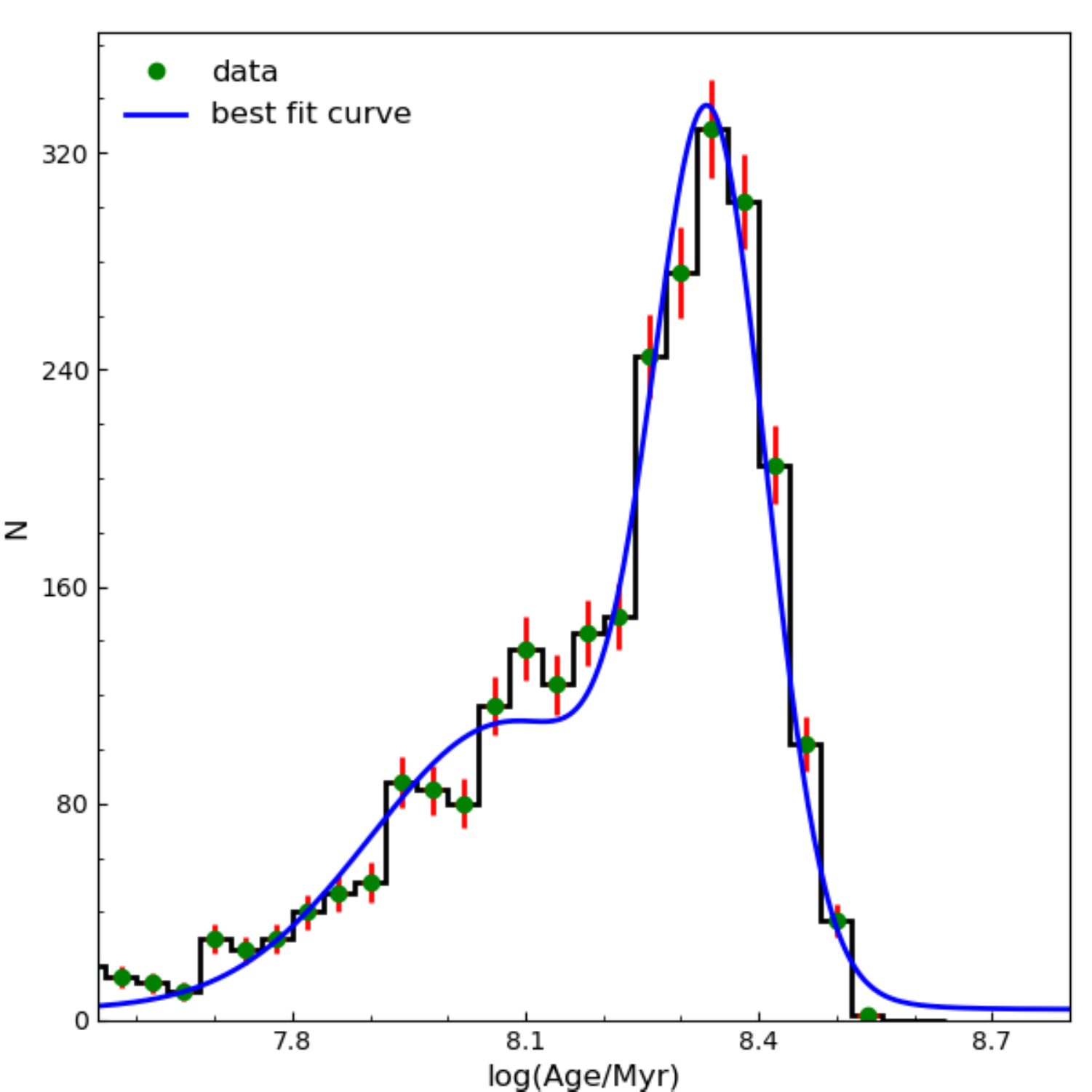}
%\vspace{-2.0cm}
\caption{Age distribution of Cepheids in the SMC. The best-fit bi-modal Gaussian profile is drawn by a thick continuous line.}
\label{fig04}
\end{figure}

Like LMC, we also determined the age of each Cepheid variable in the SMC using the theoretical period-age relation given in equation~\ref{eqno2}. The age distribution of Cepheids was examined in a logarithmic bin width of 0.04 dex and drawn in Figure~\ref{fig04} where a prominent peak is quite evident with some broad features towards the left side of the peak. The error bars in Figure~\ref{fig04} are derived in a similar fashion as mentioned in section~\ref{age_l}. The distribution shows a slow increase in the age distribution followed by a sharp decrease towards older Cepheids but the increase in age distribution is still sharper than LMC Cepheids age distribution. Similar to LMC, we tested both single and double Gaussian fit for SMC Cepheids log(Age) distribution. The double Gaussian distribution peaks at $\log ({\rm t/yr}) = 8.08\pm0.12$ and $8.34\pm0.11$. These two peaks corresponds to $120^{+38}_{-29}$ and $219^{+63}_{-49}$ Myr having second peak as more prominent. The uncertainty in the peak age represents a combined error from the statistical error in the Gaussian fit and the errors on the coefficients in equation~\ref{eqno2}. Both the $\chi^2$ and K-S test statistics are mentioned in Table~\ref{fit_test} and favours double Gaussian fit. \citet{2017MNRAS.472..808R} also found a bi-modal distribution at $120\pm10$ and $220\pm10$ Myr with a promising peak at the later epoch.
%
%---------------------------- Fig04 -----------------------------------
%------------------------------------------------------------------------
%
%
\subsection{Spatial distribution}\label{spatial_s}
\begin{figure}
\centering
\vspace{-0.9cm}
\includegraphics[width=9.5cm, height=8.9cm]{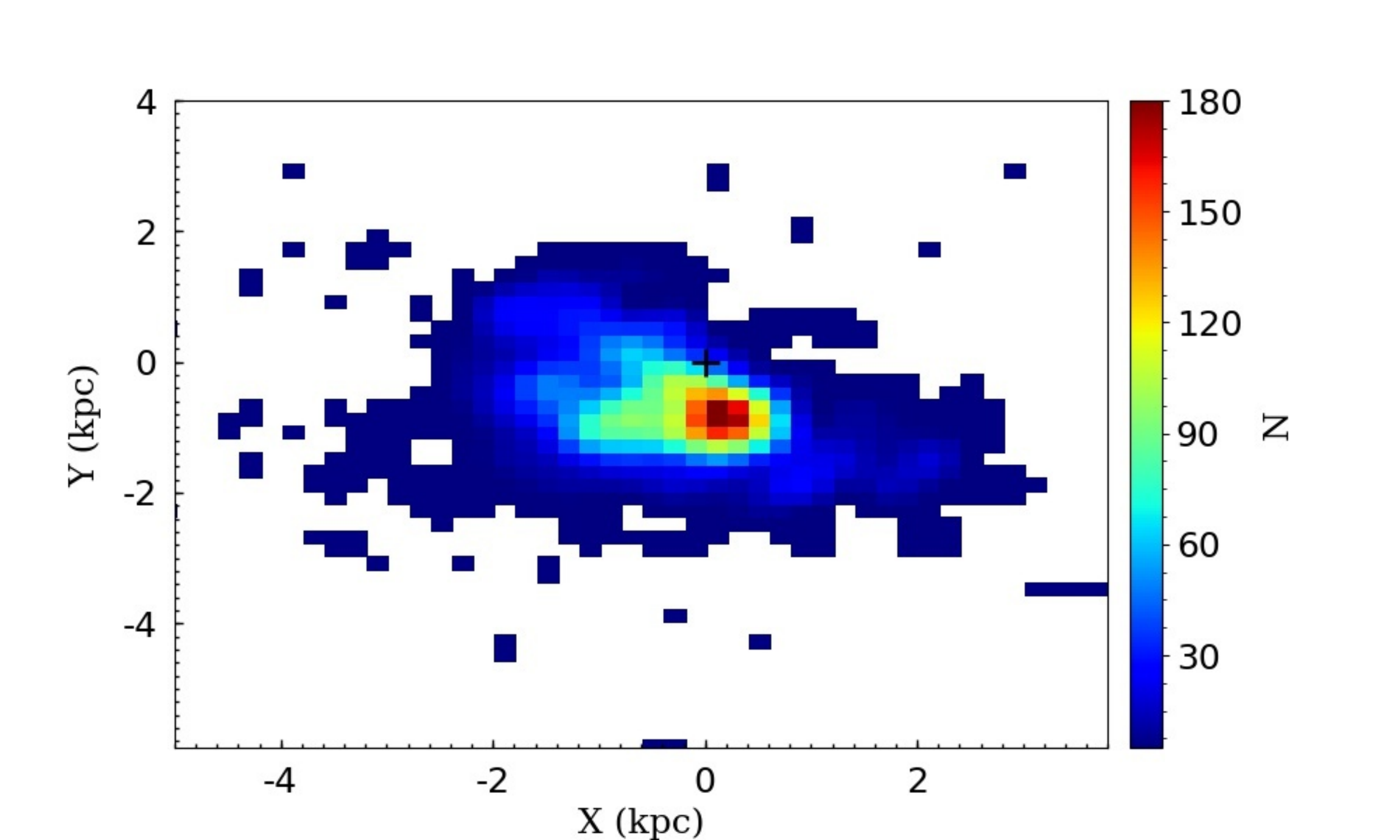}
\caption{Same as Figure~\ref{fig02} but for the SMC.}
\label{fig05}
\end{figure}

To probe the star formation in the SMC during the epoch of the Cepheids evolutionary time scale, we analyse the spatial distribution of Cepheids within the SMC. For this purpose, we divided the SMC into $90\times90$ segments with a smaller dimension of $0.2\times0.2$ kpc square, obtaining a mean spatial resolution of $\approx$ 0.22 deg$^2$. A smaller size is chosen in the case of the SMC because it contains a higher density of Cepheids per unit area in comparison to the LMC. We kept the same spatial cell size for all subsequent analyses of our study. We thus found 182 segments in the SMC. The total number of Cepheids was counted in each cell to construct the spatial map of Cepheid distribution. In Figure~\ref{fig05}, we illustrate the spatial map of Cepheids in the SMC where its known center at (X,Y) = (0,0) that corresponds to ($\alpha, \delta$) = ($00^h52^m45^s, -72^0 49^{'} 43^{''}.2)$), is marked by a plus symbol. The disk in the SMC is irregular and shows a complex geometric distribution as also noticed by \citet{2015A&A...573A.135S}. The high-density region in the map lies in the vicinity of $(X,Y) = (0.02,-0.38)$ which is towards the west of the SMC center. Such an elongation in the SMC disk releals its non-planer structure as also evident in the study of \cite{2017MNRAS.472..808R} based on the study of 717 classical Cepheids from the VISTA near-infrared survey of the Magellanic Clouds.
%
%---------------------------- Fig05 -----------------------------------
\subsection{Spatio-temporal distribution}\label{temporal_s}
\begin{figure}
\centering
\vspace{-0.9cm}
\includegraphics[width=9.5cm, height=8cm]{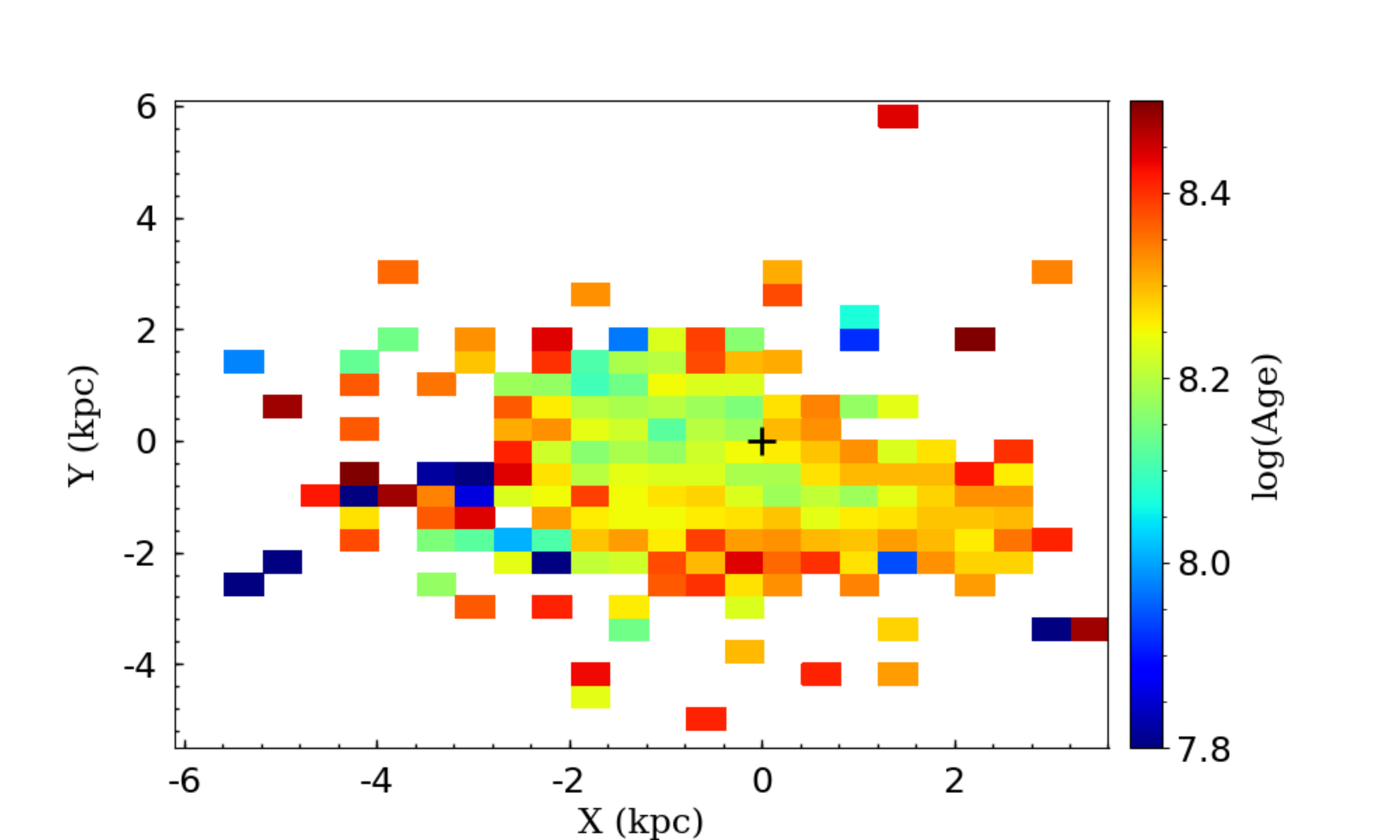}
\caption{Same as Figure~\ref{fig03} but for the SMC.}
\label{fig06}
\end{figure}

In Figure~\ref{fig06}, we draw the heat map of spatio-temporal distribution of SMC Cepheids in 182 chosen segments as discussed in section~\ref{spatial_s}. From the examination of spatio-temporal map of the SMC Cepheids, a complex and patchy age distribution is quite evident. The map shows a systematic distribution of intermediate-age Cepheids in the inner region of the SMC while older clusters as well as a few isolated young clusters are distributed in small structures towards peripheral regions. This suggests an inwards quenching of star formation in the SMC. \citet{2016AcA....66..149J} and \citet{2017MNRAS.472..808R} also found that young and old Cepheids have different geometric distributions in the SMC. They observed that closer Cepheids are preferentially distributed in the eastern regions of the SMC which are off-centered in the direction of the LMC owing to tidal interaction between the two dwarf galaxies \citep{2018MNRAS.473.3131M}.

%
%---------------------------- Fig06 -----------------------------------
%------------------------------------------------------------------------
%
\section{Discussion}\label{discuss}
The OGLE-IV survey data provided us an opportunity to statistically analyse the period and age distribution of Cepheids to probe the recent star formation scenario in the Magellanic Clouds. Taking advantage of a large and homogeneous sample of Cepheids from their precise period determination in the OGLE data, we studied the age distribution of these Cepheids from the previously reported period-age relations. The maxima in age distributions of Cepheids indicate a rapid enhancement of Cepheid formation at around $158^{+46}_{-35}$ Myr for the LMC and $219^{+63}_{-49}$ Myr for the SMC. Both the values are consistent within 1-$\sigma$. This suggests a rapid enhancement of Cepheids around 200 Myr ago in both the components of the Magellanic Clouds. This indicates a major star formation trigger at that epoch in the Magellanic Clouds. It is believed that a close encounter might have happened either between these two Magellanic Clouds components or their combined interactions with the Milky Way Galaxy that triggered an enhanced star formation rate in these two dwarf clouds \citep{2019ApJ...874...78Z}. As LMC and SMC are tidally locked, there are frequent interactions between these two components which leads to such episodic enhanced star formations in these cloud components.

It is believed that many star formation episodes have had taken place in Magellanic Clod at several epochs ranging from a few gigayears to a few million years ago \citep[e.g.,][]{2004AJ....127.1531H, 2014MNRAS.445.2214R, 2018MNRAS.478.5017R} which are often explained by the triggered mechanism due to close encounter between the two segments of the Magellanic Clouds. For example, models proposed by \citet{2006ApJ...652.1213K, 2012MNRAS.421.2109B, 2012ApJ...750...36D} and others suggest a close encounter between the SMC and the LMC that has occurred around 100-300\,Myrs ago resulting in triggered enhancement of the star formation activity in both clouds. Using the open star clusters, \citet{2010A&A...517A..50G} has studied the star formation history in the Magellanic clouds and found various episodes of enhanced star formations at 800\,Myr and 125\,Myr ago which is in accordance with the present study. The repeated tidal interaction between these two clouds thus leads to the episodic star formation events in both the dwarf galaxies and increase or decrease in star formation rates depending upon whether these two Magellanic Clouds are approaching or receding \citep{2019A&A...628A..51J}.

It is suspected that the Magellanic Bridge and/or Magellanic Stream had formed due to such frequent interactions in the past between the two components of the Magellanic Clouds \citep{2010A&A...517A..50G, 2012ApJ...750...36D, 2018MNRAS.473.3131M}. As a result of these interactions, the tidal stripping of stars and gas from the gaseous disk took place in the Magellanic system. In fact, gas in the Magellanic Bridge is believed to have been largely stripped from the SMC as a result of such interactions \citep{2014MNRAS.442.1680D, 2017MNRAS.472.2975M}. Through the observations of the intermediate-age red clump stars using deep, near-infrared photometric data obtained with the VISTA Survey of the Magellanic Clouds (VMC), \citet{2017MNRAS.467.2980S} also found strong observational evidence of the formation of the Magellanic Bridge from tidally stripped material from the SMC.
\section{Summary}\label{conc}
In this work, we studied the spatial and age distribution of 2476 Fundamental-mode Cepheids in the LMC and 2753 Fundamental-mode Cepheids in the SMC reported by the OGLE-IV survey in order to understand the star formation scenario in the Magellanic Clouds. We explicitly used Fundamental-mode Cepheids in our work as their age distribution can be well estimated through their robust period-age relations while no such relations exist for higher-mode or multi-mode Cepheids. We grouped Cepheids in small segments in both the galaxies and considered 286 such segments in the LMC with a spatial resolution of 1.2 deg$^2$ and 182 segments in the SMC with a spatial resolution of 0.22 deg$^2$ where we found at least one Fundamental-mode Cepheid. From the spatial distribution of Cepheids in both LMC and SMC, it was found that the Cepheids distribution is not planer but significantly elongated. We noticed a relatively large density of Cepheids close to center of the SMC, however, the dense region of Cepheids is substantially shifted from its center in the LMC. The south-west side of the LMC disk is densely populated in comparison to the eastern side of its disk. We did not find any larger density towards the northern arm of the LMC which shows a poor spatial density. In the case of SMC, a prominent dense clump of Cepheids was found towards the west of the SMC center. From the age distributions of Cepheids, we found an enhancement of Cepheids at around $158^{+46}_{-35}$ Myr in the LMC while it is around $219^{+63}_{-49}$ Myr in case of SMC. The two peaks are close enough which indicates an enhancement of Cepheid formation close to 200 Myrs ago in the Magellanic Clouds which is believed to be due to cloud-cloud tidal interaction in the Magellanic Clouds or that between Magellanic Clouds and Milky Way galaxy during one of the pericentric passages of the Magellanic Clouds that orbits around the Milky Way.

The spatio-temporal distributions of Cepheids in the LMC and SMC suggested slightly preferential distribution where relatively older Cepheids were segregated towards the peripheral regions in both the LMC and SMC. Most of the younger Cepheids in the SMC are found to be located in the eastern part of the cloud suggesting that the eastern side of the galaxy may be relatively younger. As both components of the Magellanic Clouds have shown evidence of repeated star formation episodes ranging from a few million years to a few giga years ago, it is essential to study Magellanic Clouds as a whole with a wide variety of stellar populations.
\section*{Acknowledgments}
This publication makes use of data products from the OGLE archive.

%%Use section* for acknowledgements
%\section*{Acknowledgements}
\vspace{-1em}

%%use \balance somewhere in the left column of the last page to balance the two columns in the end page

%%References section

%%%%%%%%%%%%%%%%%%%% REFERENCES %%%%%%%%%%%%%%%%%%
\bibliographystyle{aasjournal}
\bibliography{main}
%{}
%\nocite{*}

%%%%%%%%%%%%%%%%%%%%%%%%%%%%%%%%%%%%%%%%%%%%%%%%%%

%\end{theunbibliography}

\end{document}

%% file: tab01.tex
\begin{table}
\scriptsize
\caption{Results from single and double Gaussian fits for LMC and SMC log(Age) distributions.}
\centering
\label{fit_test}
\begin{tabular}{l l l l l}
\hline
Galaxy & Gaussian & $\chi^{2}$ & KS statistics & KS p-value \\
\hline                         
LMC    & Single   & 1114.94    & 0.3030        & 0.0772     \\
LMC    & Double   &  141.46    & 0.1515        & 0.8107     \\
SMC    & Single   & 2285.86    & 0.3250        & 0.0221     \\
SMC    & Double   &  223.02    & 0.1750        & 0.5313     \\
\hline                                                    
\end{tabular}       
\end{table}